\documentclass[%
 reprint,
superscriptaddress,
 amsmath,amssymb,
 aps,
pra,
onecolumn
]{revtex4-2}

\usepackage{graphicx}
\usepackage{dcolumn}
\usepackage{bm}

\usepackage{xcolor}

\begin{document}

\preprint{APS/123-QED}

\title{Brillouin-Induced Kerr Frequency Comb in normal dispersion fiber Fabry Perot resonators}
\author{Thomas Bunel}
 \affiliation{University of Lille, CNRS, UMR 8523 PhLAM Physique des Lasers Atomes et Mol\'ecules, F-59000 Lille, France}
\author{Julien Lumeau}
 \affiliation{Aix Marseille Univ, CNRS, Centrale Marseille, Institut Fresnel, Marseille, France}
 \author{Antonin Moreau}
 \affiliation{Aix Marseille Univ, CNRS, Centrale Marseille, Institut Fresnel, Marseille, France}
 \author{Arnaud Fernandez}
 \affiliation{LAAS-CNRS, Université de Toulouse, CNRS, 7 avenue de Colonel Roche, 31031 Toulouse, France}
  \author{Olivier Llopis}
 \affiliation{LAAS-CNRS, Université de Toulouse, CNRS, 7 avenue de Colonel Roche, 31031 Toulouse, France}
  \author{Germain Bourcier}
 \affiliation{LAAS-CNRS, Université de Toulouse, CNRS, 7 avenue de Colonel Roche, 31031 Toulouse, France}
\author{Auro M. Perego}
 \affiliation{Aston Institute of Photonic Technologies, Aston University, Birmingham, B4 7ET, United Kingdom}
  \author{Matteo Conforti}
 \affiliation{University of Lille, CNRS, UMR 8523 PhLAM Physique des Lasers Atomes et Mol\'ecules, F-59000 Lille, France}
 \author{Arnaud Mussot}
 \email{arnaud.mussot@univ-lille.fr}
 \affiliation{University of Lille, CNRS, UMR 8523 PhLAM Physique des Lasers Atomes et Mol\'ecules, F-59000 Lille, France}

\date{\today}

\begin{abstract}
We report the generation of a stable, broadband frequency comb, covering more than 10 THz, using a normal dispersion fiber Fabry-Perot resonator with a high quality factor of 69 millions. This platform ensures robust and easy integration into photonic devices via FC/PC connectors, and feature quality factors comparable to those of microresonators. We demonstrate a passive mode-locking phenomenon induced by the coherent interaction of the Kerr effect and Brillouin scattering, which generates a frequency comb with a repetition rate exceeding the free spectral range of the cavity. This parametric process modulates the continuous wave (CW) pump and can then be transformed into a train of almost square-wave pulses thanks to the generation of switching waves. Our results are supported by advanced numerical simulations, and theoretical derivations that include the Brillouin effect in the Fabry-Perot configuration. The very high stable feature of this optical frequency comb lying in the GHz range is critical to several applications ranging from telecommunication, spectroscopy and advanced microwave generation.  
\end{abstract}

\maketitle

\section*{Introduction}  
High quality factor nonlinear Kerr cavities have been instrumental in generating broadband and highly stable frequency combs, with operational repetition rate ranges ranging from MHz to THz [1, 2]. These combs are crucial for a variety of applications, such as high bit rate modern telecommunications [3], high-precision spectroscopy [4], ultra-precise distance ranging [5] and low noise microwave generation [6,7], as highlighted in recent reviews [8-10]. The characteristics of optical frequency combs (OFCs) for each application depend largely on the dispersion regime of the cavity. The most widespread phenomenon in this context are bright cavity solitons [11-13], predominantly because they facilitate broadband frequency combs, even enabling self 2f-3f self referencing schemes [14]. The latter exist in the anomalous dispersion regime. In the normal dispersion region, dark and grey solitons can be excited [3, 15-19]. These solitons, forming through the connection of two switching waves (SW) [20], produce narrower but flatter comb spectra, and offer higher pump power conversion efficiency compared to bright soliton configurations. In all these setups, Stimulated Brillouin Scattering (SBS) effect is typically avoided, either by employing pulsed pumps with durations shorter than the characteristic SBS coherence time (the phonon lifetime is typically 20 ns in silica [21]), or by designing cavities with a sufficiently large free spectral range (FSR) for the SBS gain to lie just between consecutive cavity resonances, thus preventing SBS build-up. Another approach consist in stimulating this effect to exploit its high gain to generate frequency combs in resonators. In this case, CW lasers are used as pumps, and the cavities are designed so that the SBS can resonate by matching the FSR or a multiple of the FSR with the frequency offset of the SBS (typically 10 GHz in Silica [21]). This is successfully achieved in microresonators with dimensions of about a centimeter to reach a FSR of about 10-11 GHz [22-25], or in slightly longer resonators to to match with higher order resonances [26,27]. In this process, the pump wave and the SBS-induced Stokes band resonate within the cavity, enabling highly efficient energy transfer from the pump. This transfer efficiency is so high that the Stokes wave can, in turn, generate its own SBS band, ultimately resulting in a cascading SBS effect, also assisted by a four wave mixing process (FWM) to form an OFC. These frequency combs feature generally few tens of teeth [23,24], and exhibit a significant asymmetry, as the energy transfer due to SBS occurs mainly towards the red side of the spectrum. Conversely, in longer cavities where the FSR is smaller than the linewidth of the SBS gain (typically 50 MHz in Silica [21]), several cavity modes can be amplified. The result is a complex competition between modes that can be controlled using bi-chromatic pumping [28,29], or RF locking strategies [30]. Recently, SBS has been used in fiber Fabry-Perot (FFP) resonators for a different purpose. The main objective was to counteract the cavity drift induced by the pump's thermo-optic effect [31,32], resulting in ultra-stable OFCs [33-35]. These resonators [33,34, 36-39] have emerged as a promising alternative to microresonators or fiber ring cavities. They combine a high quality factor of up to tens of millions, a compact design, excellent compatibility with other photonic components in the setup thanks to FC/PC connectors, and easy dispersion adjustment thanks to the use of existing fibers. 

Whatever the cavity architecture, the only requirement to exploit SBS for OFC generation in resonators, is to obtain a spectral overlap between the SBS gain  ($\nu_B=$ 10 GHz, $\Delta\nu = $ 50 MHz [21]), and the cavity resonances. This condition is automatically verified in fiber ring cavities, which have FSRs typically between 1 and 100 MHz, (length between 1 and 100 m) [28], or can be met by fine-tuning the cavity lengths in microresonators with FSRs in the 10 GHz range (length around 1cm) [22-24], or in FFP resonators with FSR in the GHz range (length between 1 cm and 10 cm) [33,34,36].

In this paper, we report and describe a new type of interaction between SBS and Kerr effect for generating ultra-broadband OFCs with a span exceeding 10 THz and a repetition rate of 10 GHz, precisely nine times the cavity's FSR, using resonators operating in the normal dispersion regime. This unexpected achievement is realized by leveraging the SBS effect when no overlap exists between the cavity resonances and the SBS gain curve. Our findings are based on experimental observations conducted in a few-centimeter-long FFP resonator pumped by a CW laser. Additionally, numerical simulations using a generalized mean-field equation, tailored for Fabry-Perot cavities and incorporating the SBS contribution, confirm these results. We provide a detailed analytical study elucidating the unexpected mechanism behind this phenomenon, highlighting the crucial role of SBS in a parametric process to achieve perfect phase matching alongside the SBS gain curve. 

\section*{Results}
\subsection{Experimental setup}
\begin{figure*}[h!]
    \includegraphics[width=1\columnwidth]{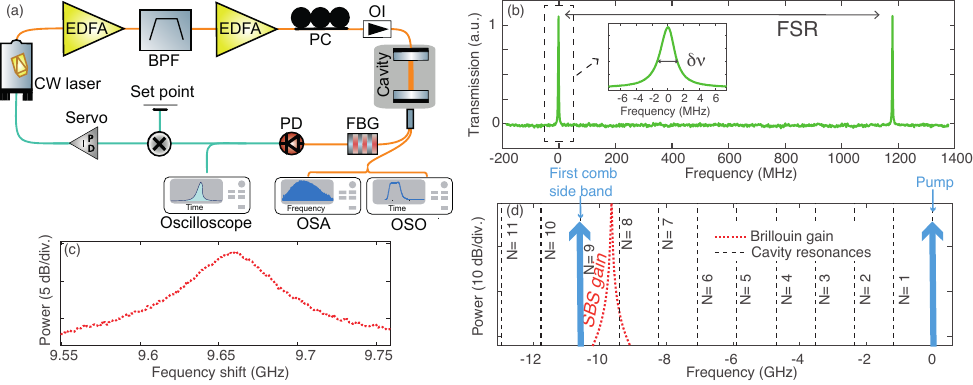}
          \caption{(a) Experimental setup. (b) Measured cavity transfer function (FSR = 1.176 GHz and $\delta\nu$ = 2.8 MHz, thus $F$=420). (c) Measured SBS gain curve (SBS shift is 9.655 GHz). (d) Cavity resonances and Brillouin gain curve. Parameters : Resonator length, $L$ = 8.78 cm, $\beta_2$ = 0.382 ps²/km, $\beta_3$ = 0.00273 ps$^3$/km and nonlinear coefficient ($\gamma$) = 10.8 /W/km. EDFA : Erbium doped Fiber Amplifier. BPF : bandpass filter. PC: polarisation controler. OI: optical isolator. FBG: fiber Bragg grating. PD: photodetector. OSA: Optical spectrum analyser. OSO: Optical sampling oscilloscope.}
       \label{fig:expsetup}
\end{figure*}

Figure \ref{fig:expsetup} (a) depicts the experimental setup. A CW laser with a central wavelength of 1550 nm is used to pump a FFP cavity of 8.75 cm length. This FFP cavity consists of a segment of a single mode normal-dispersion fiber at the pump wavelength and is connected to the rest of the setup using standard FC/PC connectors.The linear transmission function of the resonator is measured and shown in Fig. \ref{fig:expsetup}(b). The FSR of the cavity is determined to be 1.176 GHz, with a resonance's full width at half maximum (FWHM) of 2.8 MHz (see inset). This results in a finesse of 420 and a Q-factor of 69 millions. The cavity is stabilized through a feedback loop system which makes the laser pump to compensate for the cavity fluctuations. The measured gain curve of the SBS process, is displayed in Figure \ref{fig:expsetup}(c). This curve shows a central frequency shift of 9.655 GHz relative to the pump and a FWHM of around 50 MHz. It corresponds to typical values given for heavily germanium doped highly nonlinear fibers [40]. As can can be seen in  Fig. \ref{fig:expsetup} (d), the SBS gain curve is situated between two resonances of the cavity ( between $N=8$ and $N=9$), without significant overlap.

\subsection{Results and discussion}
\begin{figure}[t]
    \includegraphics[width=1\columnwidth]{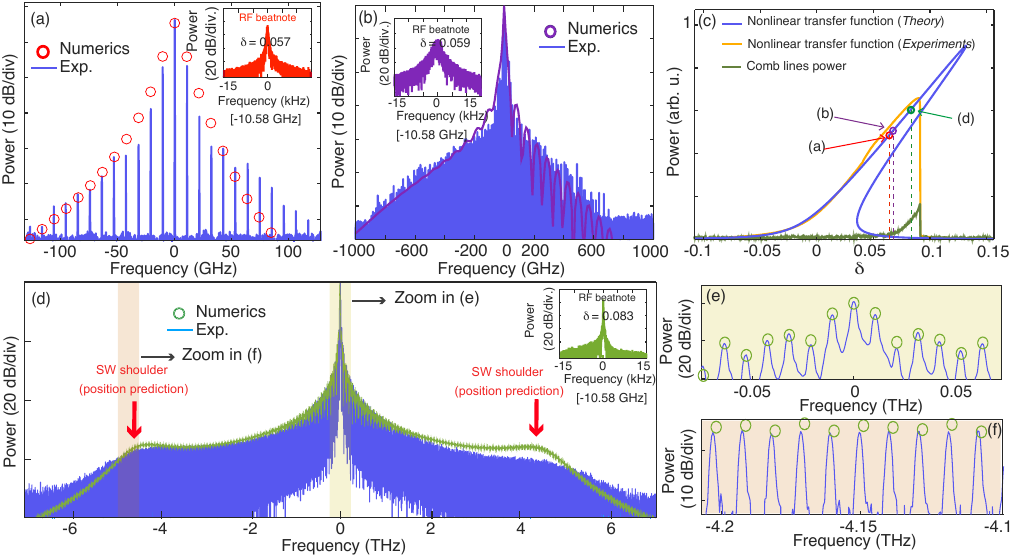}
          \caption{(a), (b) and (d) : Experimental (light blue lines) and numerical results (circles in (a) and violet and green curves in (b) and (d) corresponding to the envelope of the spectrum) for different cavity detunings ($\delta$ = 0.057, 0.059 and 0.083). The insets showcase the RF beatnote at the repetition rate (10.58 GHz) with FWHM of 1 kHz, 4 Hz and 150 Hz respectively (resolution bandwidth 10 Hz)}. (c) Nonlinear transfer function of the cavity in theory (blue curve) and experiments (yellow curve)and evolution of the comb line power (green curve). Cavity detunings used in (a), (b) and (d) are denoted by circles in red, violet and green accordingly. (e) and (f) are zooms on the spectrum shown in (d) around -4.5 THz frequency shift from the pump of (d). Vertical red arrows in (d) indicate the prediction of SW spectral shoulder [42]. The pump power at the cavity input is 0.8 Wand other parameters are listed in Methods.
       \label{fig:resultsnumexpspec}
\end{figure}

\begin{figure}[t]
    \includegraphics[width=1\columnwidth]{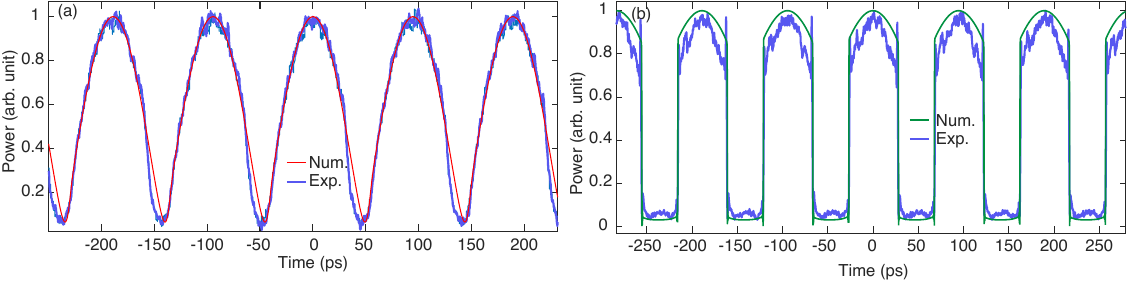}
          \caption{ (a)-(b) Time traces corresponding to Figs. 2 (a) and (d) detunings ($\delta$ = 0.057 and 0.083, respectively) measured with an OSO. Solid blue lines experiments, solid red and green lines numerics.}
       \label{fig:resultsnumexptime}
\end{figure}

The generation of OFCs in the normal dispersion regime of resonators necessitates the system operating within a bistable regime [3, 15-18].This condition enables the production of switching waves (SW), \textit{i.e} sharp fronts, propagating at a well-defined constant speed, which link the upper and lower states. The dispersion of the system tends to regularise these sharp fronts with fast oscillation at specific frequencies, and in this sense SWs may be interpreted as dispersive shock waves [41,42]. This process is not self-starting, and requires a triggering mechanism to perturb the CW pump, generating a periodic modulation that may evolve into a periodic pulse train. This early step can be achieved through various mechanisms, including mode-crossing effects [15,16], dual-pumping [28], coupled-cavity configurations [16, 17, 43], modulated pumps [18] or pulsed pumping schemes [44-48]. In the following, we will demonstrate that SBS is the triggering mechanism in our configuration.
The pump wavelength is varied from blue to red to achieve a cavity detuning ($\Delta=\delta/\alpha$, with $\alpha=\pi/F=0.0075$ the total intra-cavity losses) sufficiently large to trigger the bistable regime of the system, (reached for $\Delta>\sqrt{3}$ [2]).Subsequently, the laser is locked onto the upper branch of the cavity's response thanks to the feedback loop system. Figure \ref{fig:resultsnumexpspec} presents three distinct examples of output spectra corresponding to detunings of $\delta$ = 0.057, 0.059, and 0.083, respectively ($\Delta=$ 6.04, 8.08, and 10.74), for a pump power of 0.8 W. The  value $\delta$ = 0.083, is close to the maximum detuning  ($\delta$ = 0.086 from numerics) leading to a broad spectrum for this given pump power (see Supplemental materials). Above that limit, it is not possible anymore to excite pairs of group velocity matched SWs [42]. These values are marked by circles on the nonlinear transfer function of the cavity shown in Figure \ref{fig:resultsnumexpspec} (c). At the lowest detuning value enabling to generate a comb at this pump power ($\delta$ = 0.057), the cavity output exhibits a relatively narrow frequency comb spanning over a few tens of GHz, as depicted in Fig. \ref{fig:resultsnumexpspec} (a) (solid blue lines). A notable characteristic of this comb is the line spacing of 10.58 GHz ($N$ = =9), which does not correspond to an overlap between the SBS gain curve centered at 9.655 GHz, and a multiple of the cavity's FSR (1.176 GHz) (see Fig. \ref{fig:expsetup} (d)). We found that it is exactly nine FSRs, where practically no overlap exists between the SBS gain curve and that mode. It is crucial to note that, in contrast to the SBS comb generation in microresonators [22] where the repetition rate is determined by the largest spectral overlap between the SBS gain curve and one of the cavity resonances, our setup exhibits a different behavior. Here, as illustrated in Fig. \ref{fig:expsetup} (d), the FSR corresponds to the point of weakest overlap. The SBS gain, represented by the dashed red curve, is closer to the 8$^{\textrm{th}}$ resonance as compared to the 9$^{\textrm{th}}$, yet the repetition rate aligns with the 9$^{\textrm{th}}$ resonance, as indicated by the blue arrow. Our numerical analysis revealed that in the absence of SBS effect, the CW pump remains stable and locked on the upper state of the cavity. The SBS acts as a crucial triggering mechanism, converting the CW pump into a periodic modulation, as illustrated in Fig. \ref{fig:resultsnumexptime} (a). 
Increasing the cavity detuning to $\delta$ = 0.059 results in a broadening of the spectrum by nearly an order of magnitude, as evidenced in Fig. \ref{fig:resultsnumexpspec}(b); the line spacing consistently remains unchanged and set to 10.58 GHz. Figure \ref{fig:resultsnumexpspec}(d) illustrates that, for a larger detuning of $\delta$ = 0.083, the bandwidth can be maximally extended up to 10 THz while preserving the same repetition rate of 10.58 GHz. 
These experimental findings have been confirmed through numerical simulations based on  our novel generalisation of the Lugiato-Lefever equation (LLE) Eq. (\ref{LLE}) [49,50] (see methods). This model takes into account  the SBS contribution to the forward and backward propagating fields in the Fabry-Perot cavity, as detailed in the supplemental information. The numerical results are superimposed on the experimental data in Fig. \ref{fig:resultsnumexpspec}(a,e,f) with colored circles, and (b,d), with colored curves. Additionally, Figs. \ref{fig:resultsnumexpspec} (e) and (f) display a closer view of the central portion of Fig. \ref{fig:resultsnumexpspec} (d) and one area far from it (4.15 THz). An good agreement is observed between the experimental and numerical data. Note that the high microwave beating dynamics of about 70 dB for $\delta$ = 0.083 (inset in Fig.\ref{fig:resultsnumexpspec} (d)) compared to $\delta$ = 0.053 and 0.059 (insets in Fig.\ref{fig:resultsnumexpspec} (a) and (b)), reveal significant comb broadening for this detuning (up to 10 THz), through the generation of hundreds of strongly mode-locked frequency teeth. 

The temporal characteristics of these frequency combs were captured using an optical sampling oscilloscope (OSO) with a bandwidth of 700 GHz. For each cavity detuning we observed a stable, periodic pulse train with a repetition rate of 10.58 GHz, corresponding to a 94.5 ps period, as shown in Figs \ref{fig:resultsnumexptime}(a) and (b). As the cavity detuning increases, the leading and trailing edges of the pulses become increasingly abrupt, eventually appearing almost vertical in Fig. \ref{fig:resultsnumexptime} (b). This phenomenon is a marker of switching wave generation. Another clear indication that SWs are generated is the appearance in the spectrum of characteristic shoulders  (4.46 THz in (Fig. \ref{fig:resultsnumexpspec}(d), denoted by red arrows corresponding to the positions predicted by theoretical predictions [42]), which eventually determine the spectral extension of the comb [42]. For lower detuning values, as illustrated in Figs. Fig. \ref{fig:resultsnumexpspec}(a) and Fig. \ref{fig:resultsnumexpspec}(b), no SWs are generated. In this regime, the spectral broadening arises from the interplay of FWM and stimulated SBS, resulting in narrow and asymmetric spectra corresponding to a group velocity offset relative to the driving field (Supplemental information).
These periodic pulse trains recorded with an OSO illustrate the high stability of the combs. The agreement with numerical simulations (colored dashed lines) is almost perfect for each detuning value. To quantify the stability, we recorded the phase noise spectra  (refer to supplemental information) which shows -70 dBc/Hz at 1 kHz. The beat notes at 10.58 GHz (see the inset in Fig. \ref{fig:resultsnumexpspec} (a), (b) and (d)) are ultra narrow exhibiting 4 kHz, 1 kHz and 0.15 kHz FWHM respectively.  

\begin{figure}[tb]
    \includegraphics[width=1\columnwidth]{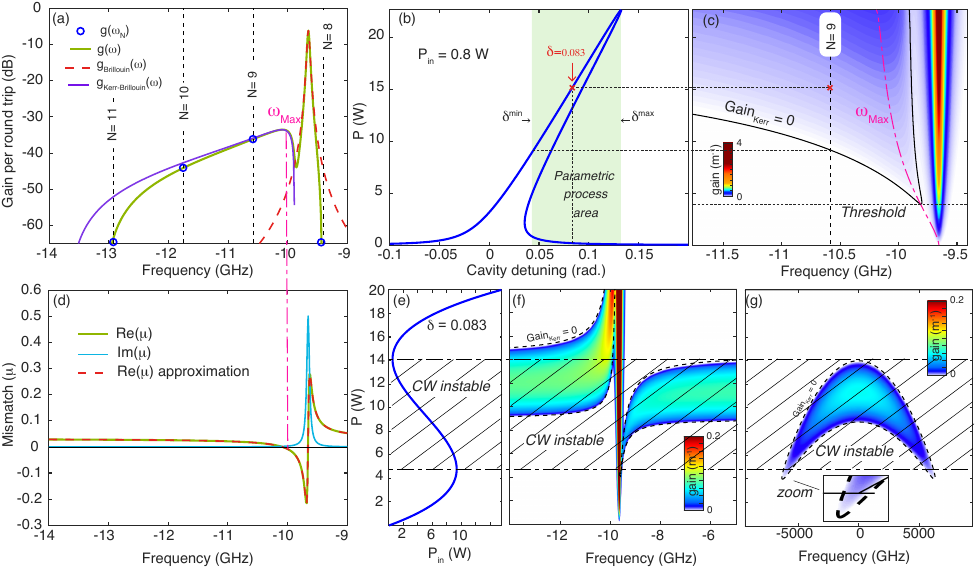}
          \caption{ (a) Parametric gain curves from theoretical predictions (green curve Eq.(\ref{gain})) for $\delta$ = 0.083 rad. The cavity resonances (vertical dashed lines) as well as the SBS gain (red dashed curve) are superimposed. The intersection between the parametric gain curve and the cavity resonances are highlighted by blue circles. The purple curve represents the  approximation of the parametric gain without the SBS gain contribution (Eqs. (\ref{gain},\ref{phase_matching})). (b) Nonlinear transfer function of the cavity for $P=0.8$ W. The green area represents the cavity detuning interval over which the parametric process can exist. (c) Parametric gain as a function of the pump power.  Black curve corresponds to the marginal stability, and the pink dashed dotted line show the evolution of the perfect phase-matching frequency (Eq. \ref{omega_max}). (d) Phase-mismatch curves (Eq. (\ref{mu})), real part (green curve), imaginary part (light blue) and approximate relation (Eq. (\ref{phase_matching}), red dashed lines). (e) Nonlinear transfer function of the cavity for $\delta$ = 0.083. (f) and (g) Parametric gain curves with and without SBS respectively for $\delta$ = 0.083. }
       \label{fig:theory}
     \end{figure}
     
To elucidate the unexpected phenomenon where the comb's repetition rate does not align with the linear cavity modes exhibiting maximum overlap with the SBS gain, we performed a linear stability analysis based on the modified LLE [Eq. (\ref{LLE})] including SBS effect (see Methods and Supplemental Information). The green curve in Fig. \ref{fig:theory}(a) represents the parametric gain spectrum described by Eq. (\ref{gain}). This curve reveals two key features: a sharp peak with the highest gain value, which corresponds to the SBS gain curve (red dashed line from
 Eq.(\ref{gain_bril})), and a broader and weaker gain band (approximately one order of magnitude lower) arising from a parametric interaction between Kerr and SBS effects. The maximum of this curve (labeled $\omega_{\text{Max}}$ in Fig. \ref{fig:theory}(a) whose expression is given by Eq. (\ref{omega_max})), results from a perfect compensation of the pump-sideband phase-mismatch via the combined action of the Kerr and SBS effects. To validate this interpretation, we remind that acousto-optic interactions in optical fibers induce both the amplification of a backward-scattered Stokes wave and a modulation of the refractive index through the electrostriction process [51]. The latter introduces a phase term, contributing to the phase-matching condition. It is highlighted in Fig. \ref{fig:theory}(d), where  we plot the approximated  phase-mismatch relation Eq.(\ref{phase_matching}), enabling to isolate the parametric process for the SBS gain one. The interaction between Kerr and SBS enables a perfect phase matching near 10 GHz (green curve in Fig. \ref{fig:theory}(d)), while eliminating the SBS contribution would give no solution, as it can be seen by the non-zero asymptotic value of the mismatch (f.i. around -14 GHz in the figure), where the Brillouin effect is vanishing, see Eq. (\ref{phase_matching}). Notably, the approximate solution (dashed red line from Eq. (\ref{phase_matching})) closely matches the exact one (green curve from Eq. (\ref{mu})). The imaginary part of Eq. (\ref{mu}) represents the standard SBS gain profile (light blue curve). Hence, the linear stability analysis highlights two critical contributions to the process: (i) the conventional SBS gain mechanism (imaginary part of Eq. (\ref{mu})) and (ii) an additional parametric contribution arising from the interplay of Kerr and SBS-induced phase effects (real part of Eq. (\ref{mu})). This extended gain bandwidth allows overlap with several cavity resonances (blue circles, $N=8$ to $N=11$ in Fig. \ref{fig:theory}(a)). The ninth resonance ($N=9$) exhibits the largest gain, thereby selecting a frequency component that does not overlap with the standard SBS gain curve (located around $N=8$). Thus, energy from the pump is transferred to this selected frequency through this parametric process, generating signal and idler waves symmetrically located around the pump. This results in pump intensity modulation at a frequency that is an exact multiple of the cavity FSR ($N=9$ in this case). The cavity thus behaves as a resonator coherently driven by an intensity-modulated pump, eventually leading to shock wave formation, significantly broadening the pump spectrum  [18, 44-48]. It is important to note that the spectral bandwidth can be controlled by tuning the fiber dispersion. Since this parameter has a negligible impact on the parametric process, the teeth spacing remains unchanged. However, the positions of the shoulders associated with SWs can be adjusted, enabling control over the comb's spectral width. This tuning allows for the adjustment of the comb width while maintaining a constant teeth spacing of 10 GHz, thereby enhancing the power spectral density of the comb (see supplemental materials).

A deeper insight into the dynamics of the process is provided in Figs. \ref{fig:theory}(b-c) and (e-g), which illustrate the system's behavior under varying cavity detuning (Figs. \ref{fig:theory}(b-c)) and input pump power (Figs. \ref{fig:theory}(e-g)). Fig. \ref{fig:theory}(b) represents the tilted transfer function of the cavity for a pump power  $P_{in}=0.8$ W. Fig. \ref{fig:theory}(c) shows a false-color plot the parametric gain as a function of frequency for operating points on the upper branch of the tilted transfer function  calculated  from Eq. (\ref{gain}). As illustrated above, for parametric instability to arise, the gain must overlap with a cavity resonance. This condition is satisfied over a specific range of cavity detuning. The minimum value (labeled $\delta^{min}$ in Fig. \ref{fig:theory}(b)) is defined by the intersection between the marginal stability curve (black curve in Fig. \ref{fig:theory}(c)) and one cavity resonance (dashed black line). The maximum detuning coincides with the tip of the tilted resonance (labeled $\delta_{max}$ in Fig. \ref{fig:theory}(b)).  
This highlights the necessity of operating with large enough detuning values to achieve a sufficiently high intra-cavity power. 
Fig. \ref{fig:theory}(e) shows the bistable function for a cavity detuning set to $\delta=0.083$. It is well-established that the upper and lower branches of the bistability curve can exhibit modulational instability depending on the cavity parameters, whereas the negative slope region cannot support any steady-state solutions [52]. This forbidden region is highlighted in Fig. \ref{fig:theory}(e-g) by the dashed area. In the standard case of normal dispersion without SBS, although the linear stability analysis predicts the existence of parametric gain in this region, the system cannot be operated in this zone and instability is observable on the lower branch only [53], as shown in Fig. \ref{fig:theory}(g). When SBS is included, the parametric gain is significantly modified around the SBS gain band (approximately 10 GHz). This modification enables the existence of gain on the accessible upper branch, as depicted in Fig. \ref{fig:theory}(f) (for better presentation the unaffected portions of the parametric far from the SBS band are not shown).

\begin{figure}[t]
    \includegraphics[width=1\columnwidth]{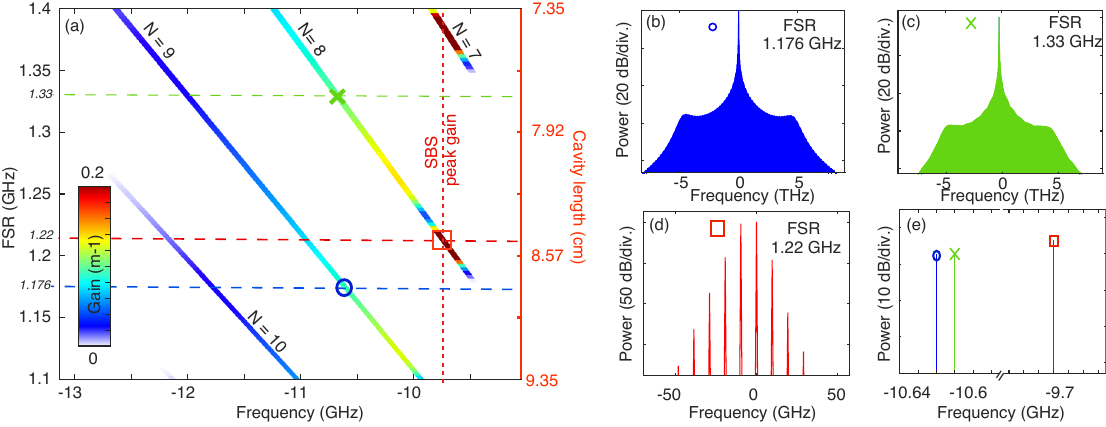}
          \caption{(a) 2D plot illustrating the evolution of the parametric gain as a function of the FSR of the cavity and the corresponding repetition rate of the frequency comb. (b)-(d) Typical examples of output spectra corresponding to a FSR of 1.176 GHz (b), 1.22 GHz (c) and 1.3 GHz (d) leading to frequency combs of 9.66 GHz, 10.4 GHz and 10.6 GHz repetition rates respectively (see (e) representing the first band from numerics corresponding to these cases). }
       \label{fig:FSRevol}
\end{figure}

To go beyond the specific configuration presented here, we plotted the evolution of the parametric gain value at several cavity resonances for various FSR values ranging from 1.1 to 1.4 GHz in Fig. \ref{fig:FSRevol}(a) (also refer to the video in additional information). This scenario  physically corresponds, for instance, to tuning the cavity length from 9.35 and 7.35 cm (see right vertical axis).As the FSR increases, the frequencies of the $N^{\textrm{th}}$ modes are shifted to larger absolute values (tilted colored lines). For clarity, we focus on the evolution around the 8$^{\textrm{th}}$ and 9$^{\textrm{th}}$ modes. At a FSR of 1.176 GHz, which matches with our experimental setup (indicated by a blue circle in Fig. \ref{fig:FSRevol}(a) and Fig. \ref{fig:resultsnumexpspec}(a)) for $\delta$ = 0.083, the parametric gain experienced by the 9$^{\textrm{th}}$ mode surpasses its neighbors, leading to the predominant transfer of pump energy to this frequency (10.58 GHz, blue circle in Fig. \ref{fig:FSRevol}(a)). The corresponding output spectrum from numerical simulations is reminded in Fig. \ref{fig:FSRevol}(b). By increasing the FSR, the parametric gains for the 9$^{\textrm{th}}$ and 10$^{\textrm{th}}$ modes decreases, while the gain of 8$^{\textrm{th}}$ mode abruptly increases due to an overlap with the SBS gain curve. This leads to a mode hopping from the 9$^{\textrm{th}}$ to the 8$^{\textrm{th}}$ mode when the latter's gain becomes the highest. As an example, at a FSR of 1.22 GHz (red square, Fig. \ref{fig:FSRevol} (a)), the output spectrum, depicted in Fig. \ref{fig:FSRevol} (d), is extremely narrow (approximately 100 GHz), comprising only a few comb teeth, similar to SBS-assisted cascading in resonators as reported in [22-24]. The very high efficiency of the SBS process, with the SBS gain being two orders of magnitude larger than the parametric gain (Fig. \ref{fig:theory} (a)), results in rapid pump depletion and almost complete energy transfer to the first Stokes sidebands. 
This process may cascade, generating additional Stokes orders, but it does not favor FWM processes due to the significantly depleted pump. The lack of power in the central part of the spectrum leads to an asymmetric spectrum. Only a few anti-Stokes lines are generated because their power and thus their number depends on the on the pump power [54] that is weak here due to to the SBS depletion. A further increase in FSR, moving further away from the SBS gain curve, allows the parametric gain to increase sufficiently to again initiate phase-matched SBS-assisted combs. An example at a FSR of 1.3 GHz (green cross in Fig. \ref{fig:FSRevol}(a)) is shown in Fig. \ref{fig:FSRevol}(c). The spectrum is almost identical to the one for an FSR of 1.176 GHz (blue circle). 
A further increase of FSR makes the 7$^{\textrm{th}}$ mode overlaps with the SBS gain curve, reproducing the behavior of a SBS-assisted cascading process similarly as for an FSR of 1.22 GHz. 
In these examples, as the FSR varies from 1.15 GHz to 1.4 GHz (corresponding to cavity lengths varying from 8.8 cm to 7.29 cm, respectively), the repetition rates of the comb varies between approximately 9.5 GHz to 11 GHz. For larger FSR values ($>$1.4 GHz, above the limits of Fig. \ref{fig:FSRevol}(e)), the frequency difference between successive cavity resonances increases, leading to less frequent mode hopping. Conversely, for  FSRs values much lower than 1.15 GHz, two cavity modes can simultaneously fall into the SBS gain band and experience similar gain values. This proximity combined with high gains can lead to competition between modes [28-30], resulting in more complex non-linear dynamics that is outside the scope of our work.


\subsection{Conclusion} 
We unveiled an original mode-locking phenomenon for generating a stable Brillouin-Kerr frequency comb in high quality factor fiber Fabry-Perot resonators. A notable aspect of our work is the discovery that the repetition rate of the OFC is not determined by the mode experiencing the maximum SBS gain, despite SBS plays a pivotal role in OFC formation. By leveraging this phenomenon, we demonstrate the generation of ultra-broadband and stable OFC spanning over 10 THz with a 10 GHz repetition rate (nine times the cavity FSR). Experimental observations in few-centimeter-length FFP resonators, under various cavity detunings, elucidate the dynamics of the process. Furthermore, our investigations have uncovered that SBS acts as an effective triggering mechanism for the generation of SW under CW pumping conditions. These findings are corroborated through advanced numerical simulations using an extended Lugiato-Lefever equation, tailored for FFP cavities and incorporating the SBS effect. Our detailed analytical study unveils the contribution of SBS in a parametric process, enabling a perfect compensation of the phase-mismatch between the pump and the generated sidebands. This breakthrough enables the generation of phase-locked stable Brillouin Kerr frequency combs, which hold significant promise for several applications, including from telecommunication, spectroscopy and advanced microwave generation.

\section*{Bibliography}
[1]	T. J. Kippenberg, A. L. Gaeta, M. Lipson, and M. L. Gorodetsky, Dissipative Kerr solitons in optical microresonators, Science 361, eaan8083 (2018).
\newline
[2]	A. Pasquazi, M. Peccianti, L. Razzari, D. J. Moss, S. Coen, M. Erkintalo, Y. K. Chembo, T. Hansson, S. Wabnitz,P. Del’Haye, X. Xue, A. M. Weiner, and R. Morandotti, Micro-combs: A novel generation of optical sources, Physics Reports 729, 1 (2018).
\newline
[3]	A. Fu¨lo¨p, M. Mazur, A. Lorences-Riesgo, B. Helgason, P.-H. Wang, Y. Xuan, D. E. Leaird, M. Qi, P. A. Andrekson,A. M. Weiner, and V. Torres-Company, High-order coherent communications using mode-locked dark-pulse Kerr combs from microresonators, Nature Communications 9, 1598 (2018).
\newline
[4]	M.-G. Suh, Q.-F. Yang, K. Y. Yang, X. Yi, and K. J. Vahala, Microresonator soliton dual-comb spectroscopy, Science 354, 600 (2016).
\newline
[5]	J. Riemensberger, A. Lukashchuk, M. Karpov, W. Weng, E. Lucas, J. Liu, and T. J. Kippenberg, Massively parallel coherent laser ranging using a soliton microcomb, Nature 581, 164 (2020).
\newline
[6]	S.-W. Huang, J. Yang, M. Yu, B. H. McGuyer, D.-L. Kwong, T. Zelevinsky, and C. W. Wong, A broadband chip-scale optical frequency synthesizer at 2.7 × 10 16 relative uncertainty, Science Advances 2, e1501489 (2016).
\newline
[7]	J. Li, H. Lee, and K. J. Vahala, enMicrowave synthesizer using an on-chip Brillouin oscillator, Nature Communications 4, 2097 (2013), number: 1 Publisher: Nature Publishing Group.
\newline
[8]	T. Fortier and E. Baumann, en20 years of developments in optical frequency comb technology and applications, Commu- nications Physics 2, 1 (2019).
\newline
[9]	S. A. Diddams, K. Vahala, and T. Udem, Optical frequency combs: Coherently uniting the electromagnetic spectrum, Science 369, 3676 (2020).
\newline
[10]	Y. Sun, J. Wu, M. Tan, X. Xu, Y. Li, R. Morandotti, A. Mitchell, and D. J. Moss, Applications of optical microcombs, Advances in Optics and Photonics 15, 86 (2023).
\newline
[11]	F. Leo, S. Coen, P. Kockaert, S.-P. Gorza, P. Emplit, and M. Haelterman, Temporal cavity solitons in one-dimensional Kerr media as bits in an all-optical buffer, Nature Photonics 4, 471 (2010).
\newline
[12]	T. Herr, V. Brasch, J. D. Jost, C. Y. Wang, N. M. Kondratiev, M. L. Gorodetsky, and T. J. Kippenberg, Temporal solitons in optical microresonators, Nature Photonics 8, 145 (2014).
\newline
[13]	N. Englebert, C. Mas Arab´ı, P. Parra-Rivas, S.-P. Gorza, and F. Leo, Temporal solitons in a coherently driven active resonator, Nature Photonics 15, 536 (2021).
\newline
[14]	V. Brasch, E. Lucas, J. D. Jost, M. Geiselmann, and T. J. Kippenberg, enSelf-referenced photonic chip soliton Kerr frequency comb, Light: Science  Applications 6, e16202 (2017).
\newline
[1]	X. Xue, Y. Xuan, Y. Liu, P.-H. Wang, S. Chen, J. Wang, D. E. Leaird, M. Qi, and A. M. Weiner, Mode-locked dark pulse Kerr combs in normal-dispersion microresonators, Nature Photonics 9, 594 (2015).
\newline
[16]	X. Xue, M. Qi, and A. M. Weiner, Normal-dispersion microresonator Kerr frequency combs, Nanophotonics 5, 244 (2016).
\newline
[17]	Q.-X. Ji, W. Jin, L. Wu, Y. Yu, Z. Yuan, W. Zhang, M. Gao, B. Li, H. Wang, C. Xiang, J. Guo, A. Feshali, M. Paniccia,V. S. Ilchenko, A. B. Matsko, J. E. Bowers, and K. J. Vahala, Engineered zero-dispersion microcombs using CMOS-ready photonics, Optica 10, 279 (2023).
\newline
[18]	H. Liu, S.-W. Huang, W. Wang, J. Yang, M. Yu, D.-L. Kwong, P. Colman, and C. W. Wong, Stimulated generation of deterministic platicon frequency microcombs, Photonics Research 10, 1877 (2022).
\newline
[19]	S.-P. Yu, E. Lucas, J. Zang, and S. B. Papp, enA continuum of bright and dark-pulse states in a photonic-crystal resonator, Nature Communications 13, 3134 (2022), publisher: Nature Publishing Group.
\newline
[20]	P. Parra-Rivas, D. Gomila, E. Knobloch, S. Coen, and L. Gelens, Origin and stability of dark pulse Kerr combs in normal dispersion resonators, Optics Letters 41, 2402 (2016).
\newline
[21]	A. Kobyakov, M. Sauer, and D. Chowdhury, Stimulated Brillouin scattering in optical fibers, Advances in Optics and Photonics 2, 1 (2010).
\newline
[22]	Y. Bai, M. Zhang, Q. Shi, S. Ding, Y. Qin, Z. Xie, X. Jiang, and M. Xiao, Brillouin-Kerr Soliton Frequency Combs in an Optical Microresonator, Physical Review Letters 126, 063901 (2021), publisher: American Physical Society.
\newline
[23]	T. F. S. Bu¨ttner, M. Merklein, I. V. Kabakova, D. D. Hudson, D.-Y. Choi, B. Luther-Davies, S. J. Madden, and B. J. Eggleton, ENPhase-locked, chip-based, cascaded stimulated Brillouin scattering, Optica 1, 311 (2014), publisher: Optica Publishing Group.
\newline
[24]	T. F. S. Bu¨ttner, I. V. Kabakova, D. D. Hudson, R. Pant, C. G. Poulton, A. C. Judge, and B. J. Eggleton, enPhase-locking and Pulse Generation in Multi-Frequency Brillouin Oscillator via Four Wave Mixing, Scientific Reports 4, 5032 (2014), number: 1 Publisher: Nature Publishing Group.
\newline
[25]	M. Asano, Y. Takeuchi, S. K. Ozdemir, R. Ikuta, L. Yang, N. Imoto, and T. Yamamoto, ENStimulated Brillouin scattering and Brillouin-coupled four-wave-mixing in a silica microbottle resonator, Optics Express 24, 12082 (2016), publisher: Optica Publishing Group.
\newline
[26]	S. Gundavarapu, G. M. Brodnik, M. Puckett, T. Huffman, D. Bose, R. Behunin, J. Wu, T. Qiu, C. Pinho, N. Chauhan, J. Nohava, P. T. Rakich, K. D. Nelson, M. Salit, and D. J. Blumenthal, enSub-hertz fundamental linewidth photonic integrated Brillouin laser, Nature Photonics 13, 60 (2019), number: 1 Publisher: Nature Publishing Group.
\newline
[27]	H. Zhang, S. Zhang, T. Bi, G. Ghalanos, Y. Zhang, H. Yan, A. Pal, J. He, S. Pan, and P. Del Haye, Microresonator soliton frequency combs via cascaded Brillouin scattering (2023), arXiv:2312.15506 [physics].
\newline
[28]	E. Lucas, M. Deroh, and B. Kibler, Dynamic Interplay Between Kerr Combs and Brillouin Lasing in Fiber Cavities, Laser  Photonics Reviews , 2300041 (2023).
\newline
[29]	Y. Huang, Q. Li, J. Han, Z. Jia, Y. Yu, Y. Yang, J. Xiao, J. Wu, D. Zhang, Y. Huang, W. Qin, and G. Qin, ENTem- poral soliton and optical frequency comb generation in a Brillouin laser cavity, Optica 6, 1491 (2019), publisher: Optica Publishing Group.
 \newline
[30]	G. Danion, L. Frein, D. Bacquet, G. Pillet, S. Molin, L. Morvan, G. Ducournau, M. Vallet, P. Szriftgiser, and M. Alouini, ENMode-hopping suppression in long Brillouin fiber laser with non-resonant pumping, Optics Letters 41, 2362 (2016), publisher: Optica Publishing Group.
\newline
[31]	K. Nishimoto, K. Minoshima, K. Minoshima, T. Yasui, N. Kuse, and N. Kuse, ENThermal control of a Kerr microresonator soliton comb via an optical sideband, Optics Letters 47, 281 (2022), publisher: Optica Publishing Group.
\newline
[32]	S. Zhang, J. M. Silver, L. D. Bino, F. Copie, M. T. M. Woodley, G. N. Ghalanos, A. Svela, N. Moroney, and P. Del’Haye, ENSub-milliwatt-level microresonator solitons with extended access range using an auxiliary laser, Optica 6, 206 (2019), publisher: Optica Publishing Group.
\newline
[33]	M. Nie, K. Jia, Y. Xie, S. Zhu, Z. Xie, and S.-W. Huang, Synthesized spatiotemporal mode-locking and photonic flywheel in multimode mesoresonators, Nat Commun 13, 6395 (2022).
\newline
[34]	K. Jia, X. Wang, D. Kwon, J. Wang, E. Tsao, H. Liu, X. Ni, J. Guo, M. Yang, X. Jiang, J. Kim, S.-n. Zhu, Z. Xie, and S.-W. Huang, Photonic Flywheel in a Monolithic Fiber Resonator, Physical Review Letters 125, 143902 (2020).
\newline
[35]	M. Nie, J. Musgrave, K. Jia, J. Bartos, S. Zhu, Z. Xie, and S.-W. Huang, enTurnkey photonic flywheel in a microresonator- filtered laser, Nature Communications 15, 55 (2024), number: 1 Publisher: Nature Publishing Group.
\newline
[36]	Z. Xiao, T. Li, M. Cai, H. Zhang, Y. Huang, C. Li, B. Yao, K. Wu, and J. Chen, Near-zero-dispersion soliton and broadband modulational instability Kerr microcombs in anomalous dispersion, Light: Science  Applications 12, 33 (2023).
\newline
[37]	E. Obrzud, S. Lecomte, and T. Herr, Temporal solitons in microresonators driven by optical pulses, Nature Photonics 11, 600 (2017).
\newline
[38]	T. Bunel, M. Conforti, Z. Ziani, J. Lumeau, A. Moreau, A. Fernandez, O. Llopis, J. Roul, A. M. Perego, K. K. Y. Wong, and A. Mussot, Observation of modulation instability Kerr frequency combs in a fiber Fabry–P´erot resonator, Optics Letters 48, 275 (2023).
\newline
[39]	T. Bunel, M. Conforti, Z. Ziani, J. Lumeau, A. Moreau, A. Fernandez, O. Llopis, G. Bourcier, and A. Mussot, 28 THz soliton frequency comb in a continuous-wave pumped fiber Fabry–P´erot resonator, APL Photonics 9, 010804 (2024).
\newline
[40]	M. Deroh, B. Kibler, H. Maillotte, T. Sylvestre, and J.-C. Beugnot, ENLarge Brillouin gain in Germania-doped core optical fibers up to a 98mol
\newline
[41]	S. Malaguti, G. Bellanca, and S. Trillo, Dispersive wave-breaking in coherently driven passive cavities, Optics Letters 39, 2475 (2014).
\newline
[42]	C. M. L. J. Bunel, Thomas, A. Moreau, and A. Mussot, Switching waves-induced broadband kerr frequency comb in fiber fabry-perot resonators, arXiv:2402.09777 (2024).
\newline
[43]	X. Xue, Y. Xuan, P.-H. Wang, Y. Liu, D. E. Leaird, M. Qi, and A. M. Weiner, Normal-dispersion microcombs enabled by controllable mode interactions, Laser  Photonics Reviews 9, L23 (2015)
\newline
[44]	Z. Xiao, K. Wu, H. Zhang, T. Li, M. Cai, Y. Huang, and J. Chen, Modeling the Kerr Comb of a Pulse Pumped F-P Microresonator With Normal Dispersion, Journal of Lightwave Technology 41, 7408 (2023).
\newline
[45]	T. Li, K. Wu, X. Zhang, M. Cai, and J. Chen, Experimental observation of stimulated Raman scattering enabled localized structure in a normal dispersion FP resonator, Optica 10, 1389 (2023).
\newline
[46]	M. Macnaughtan, M. Erkintalo, S. Coen, S. Murdoch, and Y. Xu, Temporal characteristics of stationary switching waves in a normal dispersion pulsed-pump fiber cavity, Optics Letters 48, 4097 (2023).
\newline
[47]	Y. Xu, A. Sharples, J. Fatome, S. Coen, M. Erkintalo, and S. G. Murdoch, Frequency comb generation in a pulse-pumped normal dispersion Kerr mini-resonator, Optics Letters 46, 512 (2021).
\newline
[48]	M. H. Anderson, W. Weng, G. Lihachev, A. Tikan, J. Liu, and T. J. Kippenberg, Zero dispersion Kerr solitons in optical microresonators, Nature Communications 13, 4764 (2022).
\newline
[49]	L. A. Lugiato and R. Lefever, Spatial Dissipative Structures in Passive Optical Systems, Physical Review Letters 58, 2209 (1987).
\newline
[50]	D. C. Cole, A. Gatti, S. B. Papp, F. Prati, and L. Lugiato, Theory of Kerr frequency combs in Fabry-Perot resonators, Physical Review A 98, 013831 (2018).
\newline
[51]	G. Agrawal, EnglishNonlinear Fiber Optics, Fifth Edition, 5th ed. (Academic Press, Amsterdam, 2012).
\newline
[52]	M. Haelterman, S. Trillo, and S. Wabnitz, Dissipative modulation instability in a nonlinear dispersive ring cavity, Optics Communications 91, 401 (1992).
\newline
[53]	S. Coen and M. Haelterman, enCompetition between modulational instability and switching in optical bistability, Optics Letters 24, 80 (1999).
\newline
[54]	G. Agrawal, Nonlinear Fiber Optics (Academic Press, 2007).
\newline
[55]	Z. Ziani, T. Bunel, A. M. Perego, A. Mussot, and M. Conforti, Theory of modulation instability in Kerr Fabry-Perot resonators beyond the mean-field limit, Physical Review A 109, 013507 (2024), publisher: American Physical Society.

\section*{Methods}
{\bf{Experimental setup}}. The Fabry-Perot cavity is constructed using a commercial highly nonlinear fiber (HNLF, model HN1550) with highly reflective mirrors (99.86$\%$) deposited on the faces of the fiber FC/PC connectors. The reflectivity of these mirrors remains constant over a spectral range of 100 nm around the central wavelength of 1550 nm.The CW laser used in our experimental setup has an ultra-fine linewidth of 100 Hz (NKT Koheras). To increase its power, it was amplified by a single-stage erbium-doped fiber amplifier (EDFA), followed by a two-stage EDFA, finally reaching an output power of 1.3 W. To attenuate excess amplified spontaneous emission, a narrow bandpass filter with a bandwidth of 50 GHz was strategically placed between the amplifiers. A polarisation controller was used to align the polarisation state of the laser with one of the cavity's birefringent axes. In addition, an optical isolator was placed immediately before the cavity to prevent back reflections into the laser. The output cavity spectra were recorded using a high-resolution optical spectrum analyzer (BOSA) with a resolution of 20 MHz for the narrow spectra (Figs. 2(a) and 2(b)). For the broader spectrum shown in Fig. 2(d), a standard optical spectrum analyzer was used to overcome the bandwidth limitations of the BOSA.In the time domain, we used an optical sampling oscilloscope with a wide bandwidth of 700 GHz. Part of the output was routed to a notch filter consisting of a fiber Bragg grating with a bandwidth of 100 GHz and a maximum attenuation of 40 dB. This filter was essential to eliminate a significant portion of the pump in order to avoid saturation of the stabilization system. The stabilization system, using a proportional-integral-derivative controller, adjusted the laser frequency to maintain locking with the cavity.

{\bf Theory and numerics}. 
The evolution of the optical field inside the cavity is modeled by the following mean-field equation (see supplementary information for details on the derivation):
\begin{align}\label{LLE}
\nonumber T_r\frac{\partial \psi}{\partial t}=-\left(\alpha+i\delta\right)\psi-i2L\frac{\beta_2}{2\beta_1^2}\frac{\partial^2 \psi}{\partial z^2}+\theta E_{in}&+i2L\gamma\left(|\psi|^2+X\langle|\psi|^2\rangle\right)\psi \\
+i2L&\frac{g_B}{2A_{eff}}\langle\psi\rangle\left[\langle\psi\varphi^*\rangle
+\langle\psi^*\rangle\left(\varphi-\langle\varphi\rangle\right)\right],
\end{align}
where $\psi(z,t)\; [\sqrt {\rm{W}}]$  is the slowly varying envelope of the intracavity field, $L$ the cavity length, $\beta_{1,2}$ the group velocity and group velocity dispersion of the fiber mode at the pump frequency, $T_r=2\beta_1 L$ is the roundtrip time, $E_{in}$ is the pump amplitude, $\alpha=\pi/F$ the overall losses, $F$ the cavity finesse, $\delta$ the phase detuning,  $\theta$ the mirrors' amplitude transmissivity, $\gamma \; [\rm{W}^{-1}\rm{m}^{-1}]$ the nonlinear Kerr coefficient, $X=2$ is the cross-phase modulation coefficient, $A_{eff}$ the effective area of the fiber mode, and $g_B \; [\rm{W}^{-1} m]$, $\tau_B=1/\Gamma_B=1/(2 \pi\Delta\nu_B)$  are the Brillouin gain, frequency shift and lifetime (inverse linewidth). 
We have introduced the auxiliary field $\varphi$, which can be written as the following periodic convolution 
\begin{equation}
\varphi=\psi*h_B=\int_{-L}^L\psi(\xi,t)h_B(z-\xi)d\xi.
\end{equation}
The function $h_B(z)=\sum_nh(z-n2L)$ is the periodic replication of the inverse Fourier transform (in $z$) of the Brillouin response:

\begin{equation}
H_B(k/\beta_1)=\frac{\Omega_B\Gamma_B}{\Omega_B^2-(k/\beta_1)^2-i(k/\beta_1)\Gamma_B}=\int_{-\infty}^{\infty}h(z)e^{ikz}dz
\end{equation}
\begin{equation}
h(z)=\frac{\beta_1\Omega_B\Gamma_B}{\sqrt{\Omega_B^2-(\Gamma_B/2)^2}}e^{-\frac{\Gamma_B}{2}\beta_1 z}\sin\left(\sqrt{\Omega_B^2-(\Gamma_B/2)^2}\,\beta_1z\right)\theta(z),
\end{equation}

where $\theta(z)$ is the Heaviside step function.
The angle brackets stand for spatial average $\langle.\rangle=\frac{1}{2L}\int_{-L}^L(.)dz$.

Equation (\ref{LLE}) was solved using Fourier split-step method. Linear terms were calculated exactly in frequency domain and the nonlinear terms were calculated using Runge-Kutta 4$^{th}$ method. The periodic convolutions are calculated in a straightforward manner as multiplication in the frequency domain. The parameters used for the simulations are reported in Table \ref{parameters}
\begin{table}
    \centering
    \begin{tabular}{|c|c|}
    \hline
        \textbf{Parameter} & \textbf{Value} \\
        \hline
        Cavity length, $L$ & 8.7472 cm\\
        \hline
         Free spectral range, FSR &  1.176 GHz\\ 
         \hline
        Cavity Finesse, $F=\pi/\alpha$ & 420\\
         \hline
         Mirror pwer reflectivity, $\rho^2$ & 0.998429\\
         \hline
         Inverse group velocity, $\beta_1$ & 4.86 ns/m\\
         \hline
          Group velocity dispersion, $\beta_2$ &  0.382 ps$^2$/km\\
          \hline
         Third-order dispersion, $\beta_3$ & -0.00273  ps$^3$/km \\
            \hline
     Kerre nonlinearity, $\gamma$ & 10.8 /W/km\\
           \hline
        Effective area, $A_{eff}$ & 12.4 $\mu$m$^{2}$ \\
        \hline    
        Brillouin frequency shift, $\Omega_B=2\pi\nu_B$ & $2\pi\times$9.655 GHz \\
        \hline
       Brillouin linewidth, $\Delta\nu_B=\frac{\Gamma_B}{2\pi}$ &  55 MHz \\
         \hline
          Brillouin gain, $g_B$ &  4.71$\times 10^{-12}$ m/W \\
        \hline
          $g_B/A_{eff}$ & 0.38 W$^{-1}$m$^{-1}$ \\
        \hline
    \end{tabular}
    \caption{Values of the parameters used in the numerical simulations [40] }
    \label{parameters}
\end{table}

In order to understand the physical mechanism underlying the generation of the observed SBS-Kerr combs, we performed a linear stability analysis of the constant solutions of Eq. (\ref{LLE}). The homogeneous (or CW) solutions of Eq. (\ref{LLE}) satisfy
\begin{equation}
\label{steady}\theta_1^2 P_{in}=P_s\left[\alpha^2+(\delta-2\gamma P_sL(1+X_{eff}))^2\right]
\end{equation}
where $P_s=|\psi_s|^2$ is the intracavity power  and $P_{in}=|E_{in}|^2$ .
The effective XPM coefficient is $X_{eff}=X+g_B\Gamma_B/(2A_{eff}\gamma\Omega_B)$.
We consider a perturbed cw solution $\psi(z,t)=\psi_s+\varepsilon_{+n}(t)e^{ik_nz}+\varepsilon_{-n}^*(t)e^{-ik_nz}$, where $\varepsilon_n\ll\psi_s$ and linearise around the CW state. The amplitudes of the modal perturbations obey two coupled linear ordinary differential equations $d/dt(\varepsilon_{+n},\varepsilon_{-n})^T=M_n(\varepsilon_{+n},\varepsilon_{-n})^T$. The eigenvalues of the matrix $M_n$ determine the stability of the CW solution.
 The gain of the perturbations for $n\neq0$ reads :
 \begin{equation}\label{gain}
g(\omega_n)=\frac{1}{2 L}{\rm Re}\left[-\alpha+\sqrt{(2\gamma P_s L)^2-\mu_n^2}\right],
 \end{equation}
 where 
 \begin{equation}\label{mu}
    \mu_n=-\delta+2L\frac{\beta_2}{2}\omega_n^2+2 P_s L[(2+X)\gamma+\frac{g_B}{2 A_{eff}}H_B^*(\omega_n)].
 \end{equation}
 %
 %
If we  take ($\alpha=\delta=\gamma=\beta_2=0$) in Eq. (\ref{gain}) we obtain the Brillouin gain function:
 \begin{equation}\label{gain_bril}
     g_{Br}(\omega)=-\frac{g_B}{2 A_{eff}}P_s{\rm Im}[H_B(\omega)]=-\frac{g_B}{2 A_{eff}}P_s\frac{\omega\Omega_B\Gamma_B^2}{(\Omega_B^2-\omega^2)+(\Gamma_B\omega)^2}.
 \end{equation}

 Equation (\ref{gain_bril}) represents a double Lorentzian, with maximum gain (absorption) at negative (postitive) frequency shift $\mp \Omega_B$. By expanding the  Eq. (\ref{gain_bril}) around $-\Omega_B$ we recover the usual Lorentzian shape of the Brillouin gain function [54]

It is well known that in addition to an amplification, SBS induces a modification of the effective index of the medium to respect due to Kramers-Kronig relations (causality). In order to quantify this effect, we insert the real part of the expansion of $H_B(\omega)$ around $\Omega_B$ in Eq. (\ref{phase_matching}), to obtain
 \begin{equation}\label{phase_matching}
 \mu=-\delta+2 P_s L\left[(2+X)\gamma+\frac{g_B}{2 A_{eff}}\frac{\delta\omega}{1+\delta\omega^2}\right],
 \end{equation}
where we consider $\omega_n$ as a continuous variable, neglect the dispersion and define $\delta\omega=2\frac{\Omega_B-\omega}{\Gamma_B}$. It is evident from Eq. (\ref{gain}) that the gain is maximized when $\mu=0$, which can be interpreted as a phase-matching condition. The physically relevant phase-matching frequency can be written as:
\begin{equation}\label{omega_max}
   \omega_{max}=\Omega_B+\frac{\Gamma_B}{2} \frac{g_B}{2 A_{eff}} \frac{2PL}{8\gamma P L-\delta}.
\end{equation}
Following the same procedure around $-\Omega_B$ we find the symmetric value $-\omega_{max}$, as expected from the parametric nature of this process.  The instability threshold can be found by setting $g(\omega_{max})=0$, which gives
$P_{th}=\frac{\alpha}{2\gamma L}$. This value is identical to the conventional MI threshold  in FP resonators [54].


\section*{Data availability}
Data are available upon request.
\section*{Code availability}
Codes are available upon request.
\section*{Acknowledgments}
This work was supported by the Agence Nationale de la Recherche (Programme Investissements d’Avenir, FARCO projects (AM)); Ministry of Higher Education and Research; Hauts de France Council (GPEG project, (AM)); European Regional Development Fund (Photonics for Society P4S, (AM, MC and AK)) and the CNRS (IRP LAFONI, (AM)) and H2020 Marie Skłodowska-Curie Actions (MEFISTA, MSCA-713694, (AM)), Royal Academy of Engineering (EPSRC Project EP/W002868/1, and the University of Lille Through the LAI HOLISTIC (AM).

\section*{Author contributions}
T.B and A.M. designed and performed the experiments, A. F., O. L. and G. B. fabricated the resonators, J.L. and A.M deposed the mirors on the cavities. T.B. and M.C. performed numerical simulations, M.C and A.P. performed theoretical developments. All authors contributed to analyzing the data and writing the paper. 
\section*{Competing interests}
The authors declare no competing interests.
\newpage
\end{document}